\documentclass[a4paper,12pt]{article}
\usepackage{amsmath}
\usepackage{pstricks}
\usepackage{pst-node}
\usepackage[ansinew]{inputenc}
\usepackage{amssymb,amsmath}
\usepackage{amsfonts}
\usepackage{epsfig}
\usepackage[english]{babel}
\usepackage{graphics}

\def\b{\beta}

\def\o{\omega}
\def\e{\varepsilon}

\author{Elena Kartashova\\\\
RISC, J.Kepler University, Linz, Austria\\
\\\\  e-mail: lena@risc.uni-linz.ac.at
}

\title{Kinetic equation and clipping - two limits of wave turbulence theory.\footnote{Author acknowledges support of
the Austrian Science Foundation (FWF) under projects SFB
F013/F1304.}}

\begin{document}
\date{}
\maketitle


\abstract {Different dynamics, described by kinetic equation and
clipping method is shown as well as a role of approximate
resonances in wave turbulence theory. Applications of clipping
method are sketched for gravity-capillary and drift waves. Brief
discussion of possible transition from continuous spectrum (=
kinetic equation) to discrete spectrum (= clipping) is given at
the end.}

\section{Introductory remarks}
 Wave kinetic equation has been
developed in 60-th (see, for instance, \cite{has1962},
\cite{phil61}) and applied for many different types evolution
PDEs. Kinetic equation is approximately equivalent to the initial
nonlinear PDE but has more simple form allowing direct numerical
computations of each wave amplitudes in a given domain of wave
spectrum. Wave kinetic equation is an averaged equation imposed on
a certain set of correlation functions and it is in fact one
limiting case of the quantum Bose-Einstein equation while the
Boltzman kinetic equation is its other limit.  Some statistical
assumptions have been used in order to obtain kinetic equations
and limit of its applicability then is a very complicated problem
which should be solved separately for
each specific equation \cite{zak1999}.\\

Simply formulated, conditions used for obtaining of kinetic
equation can be described as following:

\begin{itemize}
\item{} each wave takes part in resonant interactions,

\item{} each wave interacts with all other waves simultaneously,

\item{} all wave amplitudes are of the same order.
\end{itemize}

As a result, original evolution equation has been reduced to an
equation of a form
$$
\frac{d}{dT}A_i=\int \mathcal{G}(\vec{k}_i)\delta(\sum
\vec{k}_i)d\vec{k}_i
$$
which is solved in respect to each separate wave amplitude $A_i$.
Here function $\mathcal{G}$ depends on the form of initial
nonlinear PDE and notation $\delta(\sum \vec{k}_i)$ is used for
delta-function which is equal to zero on the solutions of the
system of equations describing exact resonances among interacting
waves

\begin{eqnarray}\label{res}
\begin{cases}
\omega (\vec k_1) \pm \omega (\vec k_2)\pm ... \pm \omega (\vec k_{n+1}) = 0,\\
\vec k_1 \pm \vec k_2 \pm ... \pm \vec k_{n+1} = 0.
\end{cases}
\end{eqnarray}

 and is non-zero otherwise. Wave kinetic
equation is often written out in the form
$$
\frac{d}{dT}A^2_i=\int \mathcal{G}(\vec{k}_i)\delta(\sum
\vec{k}_i)d\vec{k}_i
$$
because amplitude square is proportional to the wave energy and
this form allows to treat the results produced by kinetic equation
in terms of energy exchange between the interacting waves.\\

Existence of resonances, i.e. solutions of Sys.(\ref{res}), in a
$(n+1)$-wave system for some specific PDE allows us to replace
 the PDE by kinetic equation  which govern energy transfer through the spectrum.  A very
 interesting  fact is that even
 some classification of dispersive PDEs due to their
 integrability properties was  constructed basing on solutions of Sys.(\ref{res}). We
 present it below very briefly and
 for more details see \cite{shul}. \\

 First of all, it was proven that
in case when Eq.(\ref{res}.1) does not have any solutions, initial
nonlinear PDE could be transformed into some linear PDE by canonic
transformations.  If, on the other hand, Eq.(\ref{res}.1) does
have some solutions, then original PDE still has  nonlinearity
after canonic transformation and it can be written out as

 \begin{equation}
\Sigma_{i}  \frac{T_i \delta (\vec k_1 \pm \vec k_2 \pm ... \pm
\vec k_{i})} {\omega (\vec k_1) \pm \omega (\vec k_2)\pm ... \pm
\omega (\vec k_{i}) }\nonumber
\end{equation}

where $\delta$ is a delta-function. Vertex coefficient $T_{i_0}$
for some specific  $i_0$ is just  generalization of interaction
coefficient from the system for slowly changing amplitudes in case
of
 $i_0$-waves interactions.  In case of zero vertex coefficient, $T_i = 0$ we fall into
 the class of equations having
  solitonic solutions and in case
 of non-zero vertex coefficients, as it was shown before, the kinetic equation is constructed
 which to some extent is equivalent
  to the original PDE. \\

 Kinetic
 equations have been written for infinite
 interaction domains,
  i.e.  for an infinite plane or an infinite channel, and were used successfully for
  about 25 years  for description of many types of waves,
  mostly in cases for 3- and 4-waves interactions. The results of laboratory experiments
   showed that "reasonable
  agreement is normally obtained
  with theoretical results for the infinite case" \cite{craik}, if
   the wavelength are small  enough in comparison to the size of the experimental basin.
   The cases when wavelengths are comparable with
   the characteristic sizes of the experimental basin remain, as a rule, unexplained
   \cite{ham}   and
 have been named "effects of finite lengths" \cite{craik}.
 Attempts to put some additional physically relevant terms into the kinetic
 equations in order to make them
 applicable to long-wavelength
 systems have failed. The author of the pioneering work in  this field, O. Phillips, who
 obtained the first kinetic equation
 in \cite{phil61} wrote in
 \cite{phil85} that "new physics, new mathematics and new intuition is required"
 in order to understand  energetic
 behavior of  large-scale systems).\\

Roughly speaking, large-scale wave system is a system where wave
 do notice the boundaries. It means that
original PDE has to be regarded with zero or periodical boundary
conditions and correspondingly wave spectrum is discrete. Now
Sys.(\ref{res}) is a system of Diophantine or algebraic equations
to be solved in integers and
 assumptions used for
obtaining of kinetic equations in case of continuous wave spectrum
together with discrete Sys.(\ref{res}) have to be scrutinized
closely. Why do any pair of integers $(m,n)$ has to be part of
some solution? Or, even more, to be part of infinitely many
solutions? Is it really true that this algebraic
system  always has infinitely many solutions? Rather not.  At least, one has to prove it. \\

It was shown \cite{kar0}, \cite{kar1} that the study of discrete
systems demands  a new approach, different from method of kinetic
equation, due to their quite different properties:
\begin{itemize}
 \item{} all interacting waves are partitioned into disjoint
 classes, there is no energy flow between these classes and it is
 possible to study each class independently; the number of
 elements of a class is not large, mostly classes consist of a few
 waves only;
 \item{} there are many waves which do not take part in any resonant interactions at
 all (for instance, in case of atmospheric planetary waves this amounts to 60 $\%$ of all
 waves);
 \item{} wave interaction is local in a sense that for any fixed
 wave its interaction domain (i.e. the one which contains all the waves which
 can interact with a given one) is finite and can be written out
 explicitly;
 \item{} number of interacting waves depend drastically on the shape of a basin
 (for example, on the ratio of  its sides); there exist many basin
 where wave interactions are not possible; for some fixed
 dispersion $\o$ it is possible to describe set of all such
 basins.
 \end{itemize}

 In
particular, it means that  discrete wave systems can not be
described in terms of kinetic equation and has to be studied
separately.\\

 All the results \cite{kar1} were obtained for exact resonances
 while in \cite{kar2} it was shown that specific properties of
 discrete systems are still valid for some non-zero resonance
 width
\begin{equation} \label{width}
0< \Omega = \omega(\vec k_1) \pm \omega(\vec k_2) \pm ... \pm
\omega(\vec k_1).
\end{equation}
The existence of allowed resonance width gave rise in \cite{kar2}
to formulation of Clipping method and these results we briefly
present in Section 2, just for completeness of exposition. which
we briefly present in Section 2. Applications of Clipping method
for two physical problems are demonstrated in Sections 3 and
Section 4. Brief discussion is given in Section 5.

\section{Approximate resonances}

In order to answer the questions about resonance width, we need
some estimation of $\Omega$ as a linear form on the value of the
function $\omega$ taken in different points of its definition
domains depends on the $\omega$ changing domain.
Let us consider cases.\\

1.  Let $\omega : \mathbb{Z} \times \mathbb{Z} \rightarrow
\mathbb{Q}$, where $\mathbb{Z}$ denotes the set of integers and
$\mathbb{Q}$ denotes the set of rational numbers. Then obviously
$\Omega$ can be represented as difference of two rational numbers
$a/b$ and $c/d$ and we have trivial estimation

\begin{equation}
|\Omega |=  |\frac{a}{b}  - \frac{c}{d}| \ge \frac{1}{bd} > 0,
\end{equation}

As illustrative example for this case one may take spherical Rossby waves with $\omega=-\frac{2m}{n(n+1)}$.\\

2.  Let $\omega : \mathbb{Z} \times \mathbb{Z} \rightarrow
\mathbb{Q}(\alpha)$,  where $\alpha$ is an algebraic number, i.e.
it is a zero  of some polynomial

\begin{equation}
P(x) = a_0x^r + a_1x^{r-1} + ... + a_r,\nonumber
\end{equation}

where $a_i$ not all equal to zero. The field $\mathbb{Q}(\alpha)$
denotes the algebraic expansion of $\mathbb{Q}$, i.e., the set of
numbers of the form

\begin{equation}
 \frac{a}{b}  + \frac{c}{d} \alpha ,\nonumber
\end{equation}

with $a,b,c,d \in \mathbb{Q}$. In this case we may use
generalization of the Thue-Siegel-Roth theorem \cite{tue}: If the
algebraic numbers $\alpha_1, \alpha_2, ...,\alpha_s$ are linearly
independent with 1 over $\mathbb{Q}$,
 then for any $\epsilon>0$ we have

\begin{equation}
 |q_1 \alpha_1 + q_2 \alpha_2 + ... +q_s\alpha_s - p|> c q^{-s-\epsilon} ,\nonumber
\end{equation}

for all $p, q_1, q_2, ..., q_s \in \mathbb{Z}$ with $q = \max_i  |
q_i |$.  The constant $c$ has to be constructed for every specific
set of algebraic numbers separately.  As illustrative example
for this case one may take capillary waves with $\omega = (m^2+n^2)^{3/2}$.\\

3. Let $\omega : \mathbb{Z} \times \mathbb{Z} \rightarrow
\mathbb{R}$, where $\mathbb{R}$ denotes the set of real numbers,
$\omega$ is an arbitrary real-value function of integer variables.
To find frequency discrepancy $\Omega$ in the {\it finite}
spectral domain $D$ with some $T$ such that

\begin{equation}
D=\{(m,n): 0 < m,n \le T < \infty\},\nonumber
\end{equation}

it is enough to calculate it as \\

\begin{equation}
\Omega =min_p \Omega_p,\nonumber
\end{equation}

where

\begin{eqnarray}
 \Omega_p =\omega(\vec k^{p}_1) \pm \omega(\vec k^{p}_2) \pm ... \pm \omega(\vec k^{p}_s),\nonumber \\
 \vec k^{p}_j = (m^{p}_j, n^{p}_j), \quad \forall j=1,2,...s,\nonumber \\
 \omega(\vec k^{p}_1) \pm \omega(\vec k^{p}_2) \pm ... \pm \omega(\vec k^{p}_s) \neq 0 \quad \forall p,\nonumber
\end{eqnarray}

and $p$ is finite because the total number of wave vectors
belonging to $D$ is finite. So defined $\Omega_p$ obviously is a
non-zero number as a minimum of finite number of non-zero numbers.
As illustrative example
for this case one may take gravity water waves with  $\omega = k \tanh \alpha k$.\\

Therefore, for a wide class of dispersion functions $\omega$ it is
possible to find the low boundary for the frequency discrepancy
$\Omega$; in a {\it finite spectral domain} this boundary
exists for $arbitrary \quad \omega$.\\

Of course, in order to apply the theoretical results of wave
turbulence theory for some specific physical problem, one has also
to compute values of wave amplitudes corresponding to the regime
of weak turbulence. In other words, one has to characterize
explicitly the nonlinearity of a system under consideration, i.e.
to choose a small parameter $0<\e << 1$ describing a weakly
nonlinear regime. For instance,  for planetary waves the ratio of
the particle
  velocity to the
  phase velocity is often taken as
 the small parameter
which makes it possible to obtain \cite{kar2} following estimation
for a wave amplitude $a(m,n)$ with a wave vector $\vec{k}=(m,n)$:
$$
|a(m,n)| < \frac{6mn!2^{2n+1-m}}{5n(n+1)^{m+n+3}(n-m)(5n-m-3)}.
$$
 Usually it is a non-trivial task to compute
small parameter $\e$ symbolically, and also technical facilities
has to be taken into account (see Section 3) in order to get some
plausible results when planning some laboratory experiment.\\

 It
is important to realize that in order to observe effects predicted
by wave turbulence theory,  one has to compute together 1) allowed
resonance width,  and 2) allowed magnitudes of resonantly
interacting waves.

\section{Gravity-capillary waves}

Now that we know exact results and are able to calculate some
small but non-zero resonance width let us plan some laboratory
experiment - an experimental procedure which allows laboratory
experiments to be performed in order to observe how our theory
works in reality. Water waves are probably  the most appropriate
class of wave systems for the experimental performance because
their properties have been studied profoundly and the experimental
facilities developed are very sophisticated indeed. We regard here \\
dispersion relation for gravity-capillary waves
\begin{equation}
\omega^2(\vec k) = gk +\frac{\mu}{\nu}k^3\nonumber
\end{equation}

where $h$ is undisturbed depth, $g$ is the gravitational
acceleration, $\nu$ is the density, $\mu$ is the surface tension,
$\vec k$ is the wave vector, $k=|\vec k|$ and in c.g.s. units
$g$=981, $\mu=74$ (for water "filtered of particles with nominal
sizes larger than 0.2 $\mu m$ and with temperature about 18° C",
see \cite{ham}, p.55), $\nu=1$. The ratio $\mu/\nu$ may vary
considerably even for the same liquid, for instance, if the water
is not filtered of particles then $\mu/\nu=54$; this ratio may
also depend on
 the temperature \cite{grig}.  Therefore, first of all we have
 to choose the spectral domain of interest and then find resonantly
 interacting waves corresponding to
 the given ratio $\mu/\nu$. Let us regard spectral domain $k_x, k_y \le 30$
 which corresponds to wavelengths
 in the range (0.1, 3.0 cm) and find corresponding solutions of resonance
 conditions in the form:

 \begin{eqnarray}
\omega(\vec k_1) +\omega(\vec k_2)=\omega(\vec k_3), \nonumber\\
k_{x_1}+k_{x_2}=k_{x_3}, \quad k_{y_1}+k_{y_2}=k_{y_3}, \nonumber\\
k_{x_i},k_{y_i} \in \mathbb{N}, \quad i=1,2,3 \quad
k_{x_i},k_{y_i}\le 30.\nonumber
\end{eqnarray} \\

{\it Choice of frequencies to generate}

Let us notice that when theoretically we have to find only {\it
exact} solutions, in practice all the solutions with small enough
frequency discrepancy have also to be considered. The ratio $D$ of
frequency discrepancy to the minimal driving frequency can be
chosen as the characteristic  of the discrepancy smallness and the
value of $D$ is obviously connected with experimental precision.
Driving frequency can be controlled with precision  of order
$10^{-5}$ \cite{gol};
 therefore we may only consider as resonant triads with small
 enough $D$, say $D=10^{-6}$.  Below we present
 a few solutions (three couples of numbers in square brackets
 denote the wave
 numbers of three wave vectors
 while three numbers in round brackets mean the three corresponding
 frequencies in Hz) for different liquids:\\

$Type A$\\
$\mu/\nu=75: \quad[1,2][9,1][10,3];(8.7638, 40.4435, 49.2073)$ \\
$\mu/\nu=47: \quad[1,26][16,4][17,30];(147.0295, 75.8317, 222.8612)$ \\
$\mu/\nu=27: \quad[1,10][28,6][29,16];(30.7235, 129.5023, 160.2258)$ \\
$\mu/\nu=16: \quad[1,6][4,5][5,11];(15.5681, 16.2945, 31.8626) $\\

We present here just a few solutions from the multitude of
existing solutions in order to show that for wide variety of
liquids resonant triads do exists (the value of $\mu/\nu=75$
corresponds to clear water, 8° C; $\mu/\nu=47$: glycerine
$C_3H_5(OH)_3$, 20° C;  $\mu/\nu=27$: benzol $C_6H_6$, 60° C;
$\mu/\nu=16$: benzaldehyde
$C_6H_5CHO$ film on water, 20° C).\\

In order to demonstrate one of the most striking properties of
discrete resonant systems - namely, the existence of many
non-interacting waves - we have also to compute the frequencies of
the the waves providing big frequency discrepancy.  To find them,
it is enough to solve the system of equations

 \begin{eqnarray}
D=max_{\vec k_1, \vec k_2, \vec k_3} D_{123}, \nonumber\\
D_{123} \times min(\omega(\vec k_1),\omega(\vec k_2),\omega(\vec k_3))=\nonumber\\
=\omega(\vec k_1) +\omega(\vec k_2)-\omega(\vec k_3), \nonumber\\
k_{x_1}+k_{x_2}=k_{x_3}, \quad k_{y_1}+k_{y_2}=k_{y_3}, \nonumber\\
k_{x_i},k_{y_i} \in \mathbb{N}, \quad i=1,2,3, \quad
k_{x_i},k_{y_i}\le 30.\nonumber
\end{eqnarray}

As the solutions of this system we obtain the triads of waves with
the maximal possible ration of frequency discrepancy to the
minimal driving frequency in the given spectral domain
$k_{x_i},k_{y_i}\le 30$ and for given $\mu/\nu$ (it is enough
indeed to find triads with discrepancies larger than experimental
precision, i.e. to replace its first equation by $D >0.1$.  Below
a few triads with large
discrepancies  are presented:\\

$ Type  B$\\
$\mu/\nu=75: \quad[11,15][14,15][25,30];(112.6460, 130.0788, 337.7987) $\\
$\mu/\nu=47: \quad[14,14][15,16][29,30];(98.6504, 114.4728, 295.8396)$\\
$\mu/\nu=27: \quad[4,4][26,26][30,30];(16.2595, 186.8502, 230.8321) $\\
$\mu/\nu=16: \quad[5,5][25,25][30,30];(17.8606, 137.0759, 178.8991)$ \\

Thus now we have two types of triads: type A (exact resonant
triads) and type $B$ (triads with big discrepancies). Excitation
of the frequencies corresponding to $A$-triads provides the
possibility to observe standard periodic energy exchange within
the triad. Excitation of the frequencies corresponding to some
$B$-triad will not provide any periodic motion for similar initial
conditions (i.e. for the same experimental facilities, the same
liquid and the same
magnitudes of initial wave amplitudes). \\

{\it Choice of initial amplitudes}

 Now we have found the frequencies of waves to generate but we still have to choose initial
 values for the wave amplitudes. The resonant interaction theory deals with weak nonlinearity,
 i.e. some small parameter $0 < \epsilon << 0$ has to be chosen in order to estimate
 how big the
 initial amplitudes are allowed to be: they have to be big enough in order to eliminate
 the linear
 wave propagation but at the same time the amplitudes have to be not too large in order
 to escape turbulence.\\

  There exist  different ways of choosing $\e$, the choice is defined by the
  specifics
  of the wave system. For instance, as it was mentioned above,
  for spherical planetary waves $\e$ is chosen as a ratio of the particle
  velocity to the
  phase velocity  while for water waves usually $\e = a k$ , where $a$ is amplitude of a
 wave and $k= |\vec k| $ .  This quantity characterizes the steepness of the waves and the values
 $\epsilon=0.1$ or $0.2$ correspond to the weakly nonlinear regime. Therefore,  for arbitrary
 given wave numbers it is easy to calculate the appropriate wave
 amplitudes.\\

 {\it Dependence on basin form}

 Now we are going to behavior difference in triad's behavior in different experimental basins.
 Suppose we a rectangular experimental basin; each specific mode has the form

 \begin{equation}
 [A_{mn}(T) \cos \frac{\omega t}{2} + B_{mn}(T) \sin \frac{\omega t}{2}] \cos  \frac{m\pi x}{L_x}
 \cos  \frac{n\pi y}{L_y}\nonumber
 \end{equation}

 where $A_{mn}(T),  B_{mn}(T)$ are real and imaginary parts of slowly varying mode amplitudes; $m$ and $n$
 are integers giving the number of half-lengths in the $x-$ and $y-$directions; $L_x$ and $L_y$ are the
 sides of the experimental basin and wave vector $\vec k = (k_x, k_y)= (2\pi m/L_x, 2\pi n/L_y)$.
 First notice that for the case $L_x=L_y=L$ the dispersion relation will take form

 \begin{equation}
\omega^2(\vec k) = \frac{1}{L}gk
+\frac{1}{L^3}\frac{\mu}{\nu}k^3\nonumber
\end{equation}

which means that taken above usual form of dispersion function
corresponds to the unit square domain. The resonant solutions in a
square with side $L$ could be found as solution of resonance
conditions with $\omega (\vec k)$ written as a function of $L$.
For instance, for $\mu/\nu = 16$ in square
basin with side $L=2$ there is a resonant triad\\

[1,14][23,13][24,27]; (25.0785, 50.2490, 75.3275).\\

Notice that this triad is not resonant in the square basin with $L =1$ and vice versa,
the triad\\

[1,6][4,5][5,11]; (15.5681, 16.2945, 31.8626)\\

is resonant for $L =1$ and non-resonant for $L=2$. Therefore, even
in this very simple case when basin form does not change but only
the basin sizes, no general selecting rule exists which allows us
to find resonant triads in a given basin among the triads which
are resonant  in some other basin. Each time corresponding
modification of dispersion function has to be used. For instance,
for a rectangular basin the expression

\begin{equation}
\omega^2(\vec k) = g\frac{1}{L_xL_y}[(mL_y)^2+(nL_x)^2]^{1/2} +
\frac{1}{(L_xL_y)^2}\frac{\mu}{\nu}[(mL_y)^2+(nL_x)^2]^{3/2}\nonumber
\end{equation}

 has to be used. \\

{\it Necessary experimental facilities}

 A few wave generators must be available to make it possible to generate a few
 different wave frequencies
 simultaneously. In order to measure vertical deformation of the water surface
 from its quiescent position
 a wave gauge and a video recorder can be used. A wide variety of liquids does
 provide observable results.
 An experimental basin must have walls movable in such a way that, for instance,
 the basin form remains
 rectangular while the ratio of the basin sides changes or, in case of a circular
 basin, only the radius
 of the basin changes.

\section{Drift waves} Till now our main
interest was to investigate resonantly interacting waves because
they play significant role in the energy transfer {\it via} the
wave spectrum. But the opposite
 opposite situation - behavior of
{\it the waves which do not interact resonantly} could also be of
a great interest in some physical problems, for instance, for
drift waves in Tokamak. The following questions arise: How long do
these waves keep their energy and does the confinement time change
from one mode to another? Does there exist a way to characterize
the confinement time for every specific wave? Does this time
depend on the resonator form? etc.  Evolution of drift waves in
Tokamak plasma is described by Hasegawa-Mima equation which
coincides up to renaming of variables with barotropic vorticity
equation (BVE) which we used for numerical experiments described
below. A possibility of  energy concentration in a given set of
drift waves due to some instability mechanisms in plasma was
conjectured by V.I.Petviashvili. Basing on this conjecture, we
performed numeric simulations in which the most part of the total
initial energy of the wave field (till 60$\%$) was located within
a few specific wave groups while the rest of the energy was
distributed randomly (Gaussian) or equally among all other waves
of the spectrum. The results of numerical simulations show that
all the waves can be
partitioned into three classes having qualitatively different energetic behavior.\\

{\it Class A - active waves.} It consists of the waves taking part
in resonant interactions (RI) which are exact solutions of
resonant conditions in general form

\begin{eqnarray}
\omega(\vec k_1) \pm \omega(\vec k_2) \pm \omega(\vec k_3) =0, \label{res1}\\
\vec k_1 \pm \vec k_2 \pm \vec k_3 =0\label{res2}
\end{eqnarray}

and of the near-resonant waves which can be defined as follows.
Let us denote

\begin{equation}
R^1=(S^1_1,S^1_2,S^1_3), R^2=(S^2_1,S^2_2,S^2_3),...,
R^p=(S^p_1,S^p_2,S^p_3)
\end{equation}

the set of all wave vectors giving exact solutions of Sys.
(\ref{res1}),(\ref{res2}) in some finite spectral domain $T$ so
that $T$ contains all these wave vectors and some others which we
denote as $T^r$:

\begin{equation}
T = \{S^i_j\} \cup T^r \quad \forall i =1,2,..p, \quad
j=1,2,3\nonumber
\end{equation}

while $r=1,2,3,..., (T^2-3p^{'})$ where $p^{'} \le p$ depending on
whether or not some of the triads $R^j$ have common waves. Let us
say that  {\it waves are taking part in ARI} (approximate resonant
interactions) if they have some non-zero frequency discrepancy

\begin{equation}
0 <\omega(\vec k_1) \pm \omega(\vec k_2) \pm \omega(\vec k_3)  <<0
\end{equation}

 while  Eq.(\ref{res2}) is hold exactly. Let us consider all waves taking part
 in ARI and choose the
 triads containing {\it two waves} of some specific resonant triad $R^j$.
 Corresponding frequency
 discrepancy can be computed as

 \begin{equation}
\Omega^j= \pm \omega(S^j_i) \pm \omega(S^j_k) \pm \omega(T^r) .
\end{equation}

 Waves $T_j^r$ taking part in these interactions are called {\it near-resonant waves.}
 The class
 of active waves $A$ is now defined as follows:

\begin{equation}
A= \{S^i_j\} \cup T_l^r \subseteq T . \nonumber
\end{equation}

The wave $T_{l_0}^r$ at which the minimum of $\Omega^j$ is
achieved for some specific $j$ and $\forall T^r \in T$, is called
a {\it minimal near--resonant wave} for the triad $R^j$. Energy
leaves triad  $R^j$ {\it via} the wave $T_{l_0}^r$. Using
iteratively this procedure of the construction of the minimal
near-resonant wave for the triad $(S^j_i, S^j_k, T^r)$ and so on,
the way energy goes from one triad to another can be constructed.
Energy oscillates first among the waves within resonant triad and
then is re-distributed via a few near-resonant waves.
The {\it active waves transfer energy through the spectrum}.\\

{\it Class P - passive waves.} It consists of the waves taking
part in ARI,
 except near-resonant  waves. Therefore frequency discrepancy could be found as
$\Omega^j =  \omega(P_i) \pm \omega(P_k)\pm \omega(P_r)$ where
$P_j \in T \diagdown \{ R^p\}$ and such $p_0$ does not exist for
which some of waves $(P_i \& P_k), (P_i \& P_r), (P_k \& P_r)$
belong to resonant triad $R^{p_0}$. Passive waves conserve their
energy during time scale corresponding to the three-wave
interactions and longer. The confinement time in some specific
passive waves depends on the initial energy distribution of the
wave field. For the same initial conditions this time for
different passive waves may vary 30-60\% according
to the magnitude of discrepancy. \\

{\it Class N - neutral waves.} It consists of the waves which do
not belong to class 1 or class 2, i.e. both
Eqs.(\ref{res1}),(\ref{res2}) are not satisfied, $N= T\backslash
(A \cup P)$ and could be empty (see Table) depending on the form
of resonator. But in case of $N \neq \varnothing $ the confinement
time by these modes can be a few times longer than by passive
modes.  For illustration the graphs of the modes' energies are
drawn with a relative shift by the axes Y. \\

Numeric studies of BVE show different energetic behavior of the
modes belonging to different classes. More precisely,   energy of
the neutral mode,  is conserved during the time $T_{conserve}$
much longer then those predicted by wave turbulence theory,
sometimes 3 to 5 times longer. Energy of the passive modes, with
$\Omega = 0.21$ and with $\Omega = 0.13$ conserves  approximately
during time $T_{conserve}/2$ and afterwards begin to spread over
the wave spectrum. Energy of the active mode, $\Omega = 0$, is
changing periodically from the very beginning. At the initial
moment $T=0$ all modes have the same amount of energy.\\

 Therefore, for given dispersion relation (i.e. for given resonator form
 and boundary conditions)
 in the finite spectral domain it is possible to compute all its active,
 passive and neutral waves.
 The results  for three different forms of resonators and
 two different spectral
 truncations, T10 and T20, are following:\\

\begin{itemize}
\item{} {\bf unit sphere}
\begin{itemize}
\item  T10: Number of active modes is 4 and number of neutral
modes is 3;
 \item  T20:
 Number of active modes is 51 and number of neutral modes is 3;
\end{itemize}
 \item{} {\bf square}
\begin{itemize}
\item  T10: Number of active modes is 15 and number of neutral
modes is 0;
 \item T20:
 Number of active modes is 53 and number of neutral modes is 0;
\end{itemize}
 \item{} {\bf rectangular basin with side ratio 1/4}
 \begin{itemize}
\item  T10: Number of active modes is 4 and number of neutral
modes is 75;
 \item  T20:
 Number of active modes is 16 and number of neutral modes is 300.
\end{itemize}
\end{itemize}

 {\bf Remark. }
 The important fact is that the mode with given wave vector $\vec k = (m,n)$
 may belong to different
 classes depending on the resonator geometry. For instance, the mode with wave
 vector $(2,4)$ is
 active in a square and neutral in a rectangular basin with sides' ratio 1/4.
 Therefore, if we
 supply energy to the mode with a given frequency, the time of energy confinement
 by this mode
 is defined by the resonator geometry and we can easily increase the number of
 passive waves or
 decrease the number of  active  waves just by changing the sides' ratio of the basin.
The resonator geometry is indeed crucial fact in all these
considerations because the neutral modes do not exist
 in some geometries and therefore the best confinement time principally
 can not be obtained in them.\\

 Our main hypothesis can now be formulated as follows\\

 {\it The H-mode discharge in Tokamaks could possibly be described in terms
 of three-waves interactions of the
 drift waves. The H-modes are the neutral modes or passive modes with big
 frequency discrepancy, the
 L-modes are the passive modes with small frequency discrepancy and ELM-modes
 are the active modes.}\\

 Let us construct a few parallels between the properties of H-mode discharge
 and the properties of
 the three-waves interactions in order to support this hypothesis. We use
 the comprehensive
 review on H-mode discharges in Tokamaks \cite{grob}  to extract
 the most important facts
 on the subject.

 \begin{itemize}
 \item{} The appearance of an H-mode discharge does not depend neither on the technique for its
 obtaining nor on the Tokamak geometry.
 \end{itemize}

 - The three-wave interaction dynamics is defined completely by the initial
 energy distribution among
 the modes and therefore does not depend on way of "putting" the energy into
 the specific waves.
 The fact of this dynamics existence also does not depend on the Tokamak geometry.
 The crucial fact is
 {\it the very existence} of resonator which means the boundary conditions
 importance; as soon as modes
 have  lengths long enough "to notice" boundary conditions the dynamics
 described above will take place.
 {\it The specific geometry defines only which specific modes will be active,
 passive or neutral in
 this geometry.}

  \begin{itemize}
 \item{} The occurrence of repetitive instabilities at plasma edge - ELM-s.
 There exist ELM-s of the
 three types: "giant" (Type I), "grassy" (Type II) and Type III with big,
 small and intermediate
 frequencies correspondingly.
 \end{itemize}

- The active waves perform the periodic energy exchange between
the modes of the resonant or near-resonant triads. Numerical
simulations show that the periodic energy oscillations within the
modes of these triads take place and their frequencies belong to
the three different types described above. We do not touch here
the question about specifics of energetic behavior of the passive
waves in the presence of some active waves (it is in itself very
interesting and complicated problem) and only point out the fact
that under some conditions the presence of active waves may
decrease the confinement time of some specific passive waves.

 \begin{itemize}
 \item{} The H-mode is achieved only when a exceeded threshold in the
 heating power is exceeded
 but at the same time the improvement in core confinement is due to the
 reduction of turbulence.
 \end{itemize}

 - These high and low boundaries for the total wave field energy form, so to say,
 "the corridor" of weak
 nonlinearity: the treshold in the heating provides the transfer from the linear
 regime to a weakly
 nonlinear one while reduction of turbulence prevents transition to the strong
 nonlinear regime.

 \begin{itemize}
 \item{} The H-mode is initiated by the formation of the transport barrier;
 there exist a very strong
 temporal correlation between development of the transport barrier, a reduction
 in fluctuation of
 amplitudes and a large change in the plasma edge; the mechanism by which the
 improvement in the
 core confinement is facilitated by the transport barrier is not known.
 \end{itemize}

 - The transport barrier formation leads to some specific profiles in the flow
 velocity and, therefore,
 to the specific energy distribution among the wave spectrum. Namely, the reduction
 of turbulence in the
 most part of the plasma core and increasing turbulence at the plasma periphery
 leads to the energy
 distribution needed for the resonant interactions become dominate, i.e. the
 most part of the
 wave field
 is settled into a few modes while the rest of energy  is distributed more or
 less homogeneously among the
 others. If in an experiment we are able to increase the energy of some desired
 mode, we will obtain the
 theoretically predicted confinement.\\

 The presented analysis is neither full nor exhaustive and in any case has
 only instructive importance
 because the dispersion relations under consideration are not realistic for
 toroidal plasma. Our purpose
 here is just to attract attention to the complicated problem of finding
 realistic dispersion function(s)
 for this very interesting physical problem. This hard common work of physicists
 and mathematicians may
 be well rewarded because as a result a new Tokamak geometry may evolve in which a
 lot of waves will
 improve their energy confinement time drastically.\\

\section{Concluding remarks}

We have shown that kinetic equation and clipping describe dynamics
of two opposite limits of wave turbulence theory - continuous and
discrete wave spectra correspondingly. \\

The natural - and highly non-trivial - problem to be studied now
is to describe transition from one limit to another. One of the
most tricky point here is following. As it was noticed in Section
3, discrete systems described by the same evolution PDE, the same
eigenfunctions of linear part of original PDE and having
functionally the same dispersion relation will have different
resonant triads corresponding to slightly different boundary
conditions so that triad ([1,6][4,5][5,11]) is resonant in square
basin with a side $L=1$ (in non-dimensional units) and is not
resonant if $L=2$.\\

Barotropic vorticity equation demonstrates much more complicated
behavior, namely even the form of linear modes itself changes when
regarded with different boundary conditions. For instance, BVE on
a sphere has a linear wave
\begin{equation} \label{sphere}
\psi_{sphere} = A P_n^m (\sin \phi) \exp{i[m \lambda
+\frac{2m}{n(n+1)} t]},
\end{equation}
where  $P_n^m (x)$ is the associated Legendre function of degree
$n$ and order $m$;  dispersion relation has then form $$
\omega_{sphere} = -2m /[n(n+1)]$$
 and wave vector $\vec k = (m,n)$ has integer coordinates $m,n$.
In the case of infinite plane BVE has a linear wave has  form
\begin{equation} \label{plane}
\psi_{plane} = A \exp {i(k_{x}x+k_{y}y+\omega)}
\end{equation}
and wave vector  $\vec k = (k_{x},k_{y})$ has real coordinates $
k_{x},k_{y}$ while dispersion relation is written out as
$$\omega_{plane} =
k_{x}/(1+k_{x}+k_{y}).$$

Thus, intuitively reasonable opinion, that it is enough to regard
the resonance surface constructed for real variables and then find
(if possible) its integer points, does not hold here. Transition
from a sphere to a special infinite plane called $\b$-plane can
only be
constructed locally, at some interaction latitude and this transition
 is possible {\bf iff}
this latitude exists \cite{kar4}. In particularly, it means that
1) transition from a spherical domain to an infinite plane is in
fact transition to a one-parametric family of infinite planes; and
2) such a transition is
not always possible.\\

Nevertheless, these systems have much in common, not only general
energetic behavior but also some more fine properties. For
instance,   "corridor" of weak
 nonlinearity described in previous Section and known from
 experimental plasma researches, has been established
\cite{naz1} in numerical computations with capillary waves. A very
promising idea has been formulated in \cite{naz1} - "to explain
the slowdown of energy flux and lower Kolmogorov constant observed
by Pushkarev and Zakharov in numerical experiments \cite{zak2000}"
in terms of discrete system behavior. It is a highly demanding
task indeed to prove this hypothesis because transition from
kinetic equation to the original PDE with a given boundary
conditions will be much more complicated than just changing the
interaction domain within the frame of the same equation.

\end{document}